\documentclass[twocolumn,showkeys,aps,prb,showpacs]{revtex4-1}
\usepackage{graphicx}
\usepackage[CJKbookmarks,dvipdfm,colorlinks,linkcolor=blue,citecolor=blue]{hyperref}

\begin{document}

\title{Strain effect on power factor in  monolayer $\mathrm{MoS_2}$}

\author{San-Dong Guo}
\affiliation{Department of Physics, School of Sciences, China University of Mining and
Technology, Xuzhou 221116, Jiangsu, China}
\begin{abstract}
 Biaxial strain dependence of electronic structures and  thermoelectric properties of   monolayer $\mathrm{MoS_2}$, including  compressive and tensile  strain, are investigated by using  local-density approximation (LDA) plus spin-orbit coupling (SOC). Both LDA and LDA+SOC results show that $\mathrm{MoS_2}$ is a direct gap semiconductor with optimized lattice constants. It is found that SOC has important effect on power factor, which  can enhance one in n-type doping, but  has a obvious detrimental influence for
 p-type. Both compressive and tensile strain can induce direct-indirect gap transition, which produce remarkable influence on power factor.
 Calculated results show that  strain can induce  significantly enhanced power factor in  n-type doping by compressive strain and in p-type doping by tensile strain at the critical strain of
 direct-indirect gap transition.  These  can be explained by  strain-induced  accidental degeneracies, which leads to improved Seebeck coefficient.
Calculated results show that n-type doping can provide better power factor than p-type doping.
 These results  make us believe  that thermoelectric properties  of monolayer $\mathrm{MoS_2}$ can be improved  in n-type doping by compressive strain.

\end{abstract}


\maketitle

\section{Introduction}

Thermoelectric material by using the Seebeck effect and Peltier effect can realize hot-electricity conversion to solve energy issues.
As is well known, the efficiency of thermoelectric conversion
can be characterized by dimensionless  figure of merit\cite{s1,s2}, $ZT=S^2\sigma T/(\kappa_e+\kappa_L)$, where S, $\sigma$, T, $\kappa_e$ and $\kappa_L$ are the Seebeck coefficient, electrical conductivity, absolute  temperature, the electronic and lattice thermal conductivities, respectively.
Bismuth-tellurium systems\cite{s3,s4} and  lead chalcogenides\cite{s7,s8}  are  excellent thermoelectric material in
the application of thermoelectric devices.    Searching for high $ZT$ materials is  the main objective of  thermoelectric research,  which requires  a high electrical conductance and large Seebeck coefficient and low thermal conductance. However, they are generally coupled with each other. So, it is difficult to enhance one, but not to  adversely affect else parameter.
Low-dimensional materials have been proved to be advanced in designing high-performance thermoelectric devices, such as $\mathrm{Bi_2Te_3}$ nanowire,
monolayer phosphorene and silicene\cite{s9,s10,s11,s12,s13}.

Semiconducting  two-dimensional (2D) materials have potential application in nanoelectronics and nanophotonics. Due to the presence of  intrinsic direct band gap of 1.9 eV, monolayer
$\mathrm{MoS_2}$ have been widely investigated both experimentally and theoretically \cite{q1,q2,q3,q4,q5,q6,f1} in comparison with the gapless Graphene.
Recently, it has been applied in field effect transistors, photovoltaics and photocatalysis\cite{q7,q8,q9,q10,q11}.
 Band gap tuning is very important for electronic and photonics applications, which has been realized  by applied  strain  and electric field\cite{q12,q13,q131,q132,q14,q15}.
The thermoelectric properties related with  $\mathrm{MoS_2}$  has been widely investigated, including bulk\cite{t1}, few layers\cite{t2}, monolayers to nanotubes\cite{t3,t4} and armchair and zigzag
mono- and fewlayer  $\mathrm{MoS_2}$\cite{t5}. In these theoretical calculation, SOC is neglected, but SOC is very important for power factor calculations\cite{so1,so2,gsd3}.
The thermoelectric power factor  can be enhanced  dramatically  by applied strain\cite{gsd3,th2,th3,th4} in some thermoelectric materials.

Here, the biaxial strain dependence of electronic structures and  power factor of monolayer  $\mathrm{MoS_2}$ are calculated  by  first-principle calculations and Boltzmann transport theory , including the relativistic effect. Calculated results show that SOC can reduce
power factor in p-type doping, and that can improve one in n-type doping, which is different from the usual detrimental effects\cite{so1,so2}. It is found that  strain can induce  significantly enhanced power factor in  both n-type  and  p-type doping by tuning the electronic structures of monolayer $\mathrm{MoS_2}$, which can be explained by strain-induced  accidental degeneracies.
Compressive strain tuning can induce better power factor in n-type doping, therefore  monolayer  $\mathrm{MoS_2}$ can become more efficient for thermoelectric application in n-type doping.
\begin{figure}
  \includegraphics[width=8.0cm]{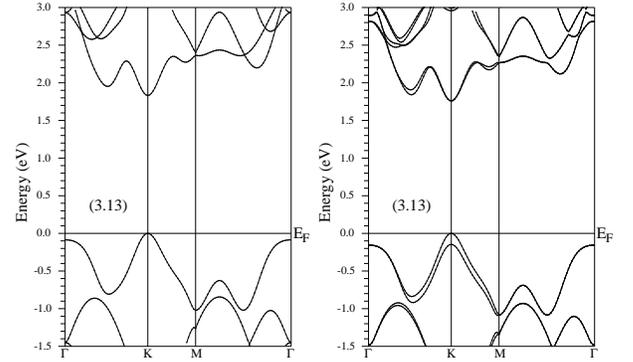}
  \caption{The energy band structures of monolayer $\mathrm{MoS_2}$ with the optimized lattice constants by using LDA (Left) and LDA+SOC (Right).}\label{t1}
\end{figure}

\begin{figure*}
  \includegraphics[width=15cm]{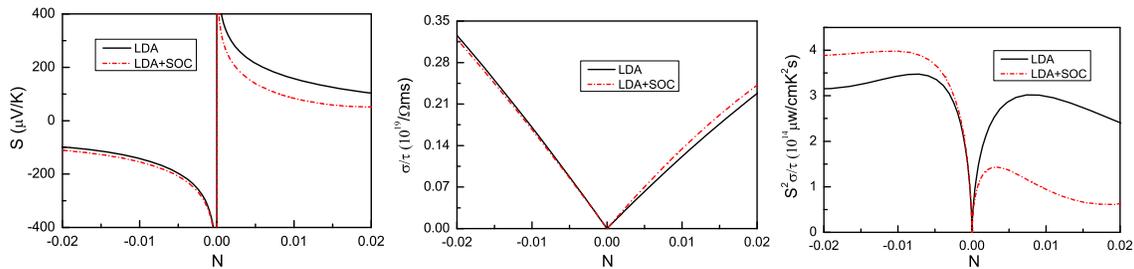}
  \caption{(Color online)  At temperature of 300 K,  transport coefficients  as a function of doping levels (electrons [minus value] or holes [positive value] per unit cell):  Seebeck coefficient S (Left), electrical conductivity with respect to scattering time  $\mathrm{\sigma/\tau}$ (Middle) and   power factor with respect to scattering time $\mathrm{S^2\sigma/\tau}$ (Right)  calculated with LDA (Black solid line) and LDA+SOC (Red dotted line).}\label{t2}
\end{figure*}

\begin{figure}
  \includegraphics[width=7.0cm]{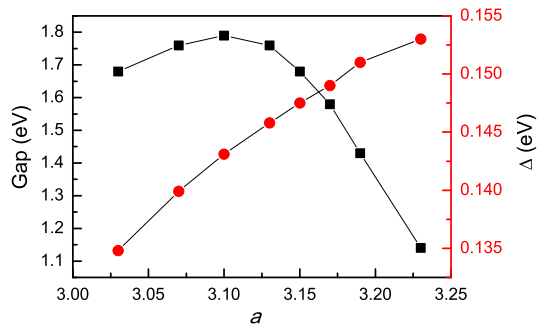}
  \caption{(Color online) The energy band gap (Gap) and  the value  of spin-orbit splitting at K point ($\Delta$) as a function of $a$ by using LDA+SOC.}\label{t3}
\end{figure}
The rest of the paper is organized as follows. In the next
section, we shall give our computational details. In the third section, we shall present our main calculated results and analysis. Finally, we shall give our conclusion in the fourth
section.

\begin{figure*}
  \includegraphics[width=12.0cm]{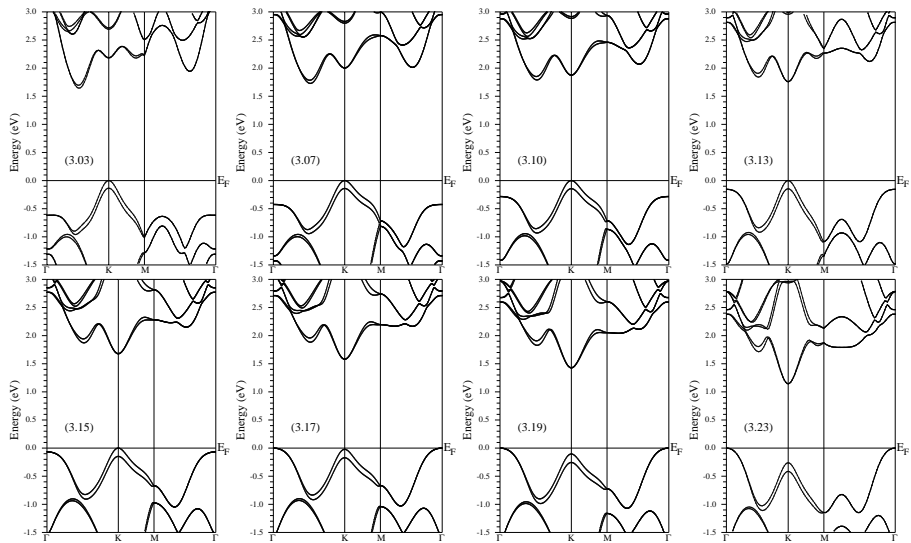}
  \caption{The energy band structures  of monolayer $\mathrm{MoS_2}$ with $a$ being from 3.03 $\mathrm{{\AA}}$ to 3.23 $\mathrm{{\AA}}$  calculated by using LDA+SOC. }\label{t4}
\end{figure*}

\begin{figure}
    \includegraphics[width=8cm]{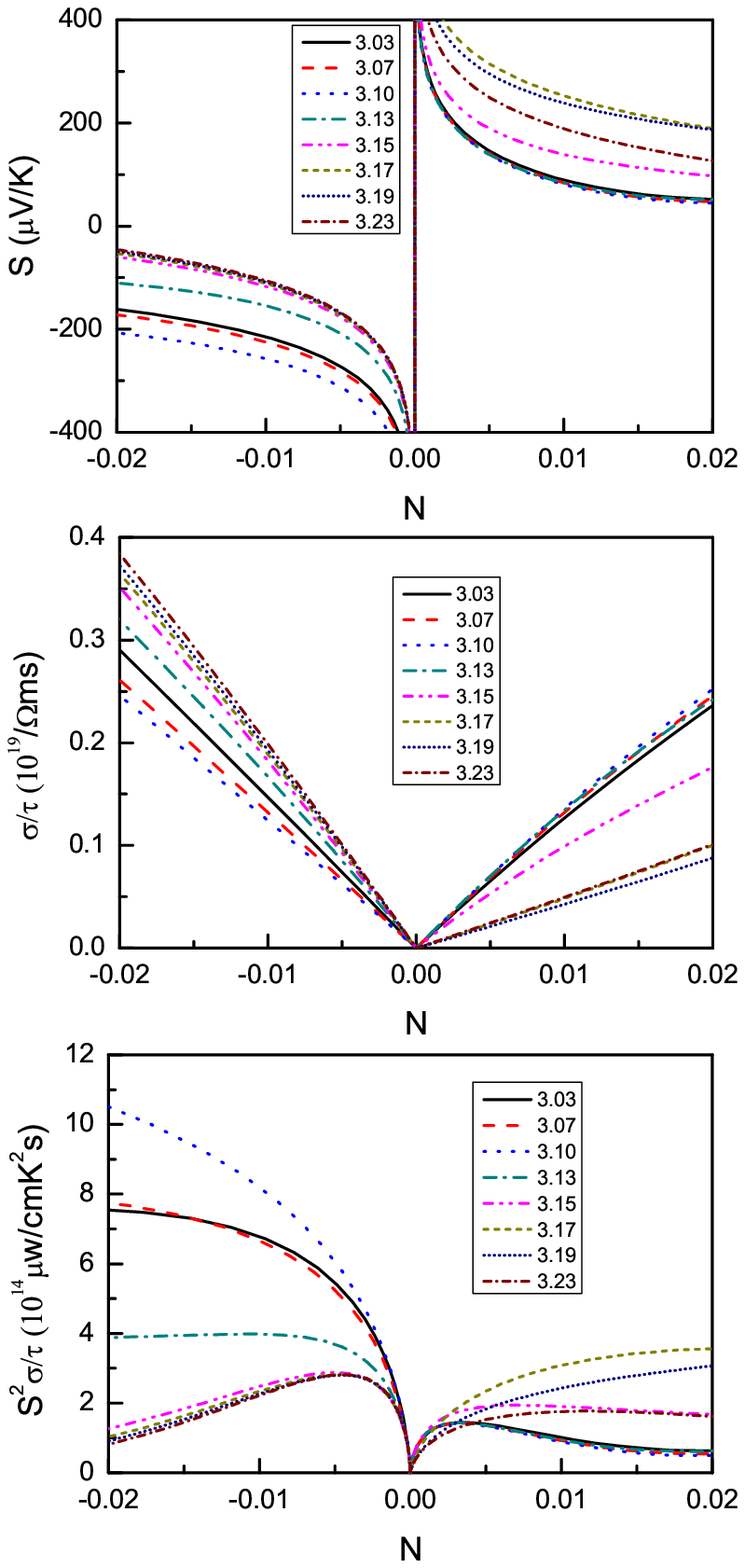}
  \caption{(Color online) At temperature of 300 K,  transport coefficients  as a function of doping levels (electrons [minus value] or holes [positive value] per unit cell):  Seebeck coefficient S (Top), electrical conductivity with respect to scattering time  $\mathrm{\sigma/\tau}$ (Middle) and   power factor with respect to scattering time $\mathrm{S^2\sigma/\tau}$ (Bottom)  with $a$ being from 3.03 $\mathrm{{\AA}}$ to 3.23 $\mathrm{{\AA}}$   calculated by using LDA+SOC. }\label{t5}
\end{figure}

\section{Computational detail}
We use a full-potential linearized augmented-plane-waves method
within the density functional theory (DFT) \cite{1}, as implemented in
the package WIEN2k \cite{2}.  We use LDA  for the
exchange-correlation potential  to do our DFT
calculations.  The full relativistic effects are calculated
with the Dirac equations for core states, and the scalar
relativistic approximation is used for valence states
\cite{10,11,12}. The SOC was included self-consistently
by solving the radial Dirac equation for the core electrons
and evaluated by the second-variation method\cite{so}. We use 6000 k-points in the
first Brillouin zone for the self-consistent calculation.
We make harmonic expansion up to $\mathrm{l_{max} =10}$ in each of the atomic spheres, and
set $\mathrm{R_{mt}*k_{max} = 8}$. The self-consistent calculations are
considered to be converged when the integration of the absolute
charge-density difference between the input and output electron
density is less than $0.0001|e|$ per formula unit, where $e$ is
the electron charge. Transport calculations
are performed through solving Boltzmann
transport equations within the constant
scattering time approximation (CSTA) as implemented in
BoltzTrap\cite{b} (Note: the parameter LPFAC can not choose the default value 5, and should choose larger value. Here, we choose LPFAC value for 20.), which has been applied successfully to several
materials\cite{b1,b2,b3}. To
obtain accurate transport coefficients, we use 200 $\times$ 200 $\times$ 1 k-point meshes in the
first Brillouin zone for the energy band calculation.

\section{MAIN CALCULATED RESULTS AND ANALYSIS}
Firstly, the crystal structure of  monolayer $\mathrm{MoS_2}$ is constructed with the
vacuum region of 20 $\mathrm{{\AA}}$ to avoid spurious interaction, and the optimized lattice constant is $a$=3.13 $\mathrm{{\AA}}$ by using LDA.
The SOC effect on electronic structures is considered, and the energy band structures  by using LDA  and LDA+SOC are plotted in \autoref{t1}.
Both LDA and LDA+SOC results show  $\mathrm{MoS_2}$ is a direct gap semiconductor, with the band gap value being 1.87 eV and 1.76 eV, respectively. Both $a$ and LDA gap are consistent with other theoretical values\cite{f1}.
It is clearly seen that SOC has obvious influence on the valence bands near  high symmetry  K point,
 leading to a spin-orbital splitting value of 0.146 eV at K point, while has a negligible effect on the conduction bands near K point. However,
 the remarkable splitting is observed  along the high symmetry $\Gamma$-K line on the conduction bands.
These SOC effects produce remarkable influence on the thermoelectric properties.

The  transport coefficients calculations, such as Seebeck coefficient S and electrical conductivity with respect to scattering time  $\mathrm{\sigma/\tau}$,  are performed  within CSTA Boltzmann theory. Any assumptions on temperature and doping
level dependence of  the band structure are not considered.  To consider SOC effects on thermoelectric properties, \autoref{t2}  shows the Seebeck coefficient S,  electrical conductivity with respect to scattering time  $\mathrm{\sigma/\tau}$ and  power factor with respect to scattering time $\mathrm{S^2\sigma/\tau}$    as  a function of doping levels  at the temperature of 300 K by using LDA and LDA+SOC. The negative doping levels  imply the
n-type doping, being related to conduction bands, with the negative Seebeck coefficient, and
the positive doping levels mean p-type doping, being connected to the valence bands, with the positive Seebeck coefficient.

 In p-type doping,  SOC has a detrimental influence on the Seebeck coefficient  S,  while has a improved effect on S (absolute value) in n-type doping.
 The opposite SOC effect on the  $\mathrm{\sigma/\tau}$ is observed for both p-type and n-type doping.
Due to the dominant role of S to power factor, the same influence of SOC on the power factor with S is found. The SOC produces larger  influence on power
factor in p-type than in n-type.  The maximum power factors (MPF) in unit of $\tau\times10^{14}$$\mathrm{\mu W/(cm K^2 s)}$ are extracted in n-type and p-type doping with LDA and LDA+SOC. The MPF by using LDA+SOC in p-type doping is about 52.5\% smaller than that with LDA, while the MPF with LDA+SOC is about 14.6 \% bigger than that by LDA in n-type.  So, the SOC has to be considered in the theoretical prediction of power factor of monolayer $\mathrm{MoS_2}$.

The effects of SOC on S can be explained by analyzing  SOC influence on the band structure. Upon opening of SOC, the conduction band extremum along the $\Gamma$-L line  moves close to  the conduction band minimum (CBM) due to the spin-orbital splitting, giving rise to more  adjacent  electron pockets (n-type), which induces higher S. However, SOC removes the  degeneracy  of valence band maximum (VBM) at the K point, reducing the slope of density of states (DOS) near the Fermi level in the valence bands (p-type), which leads to the lower S.
When SOC is included, the valence bands become
more dispersive,  increasing the mobility of p-type charge carriers, resulting in an improved  $\mathrm{\sigma/\tau}$. However, SOC has little effect on the conduction band dispersion, leading to
weak dependence of SOC on $\mathrm{\sigma/\tau}$.

Strain effect on the electronic structures of monolayer $\mathrm{MoS_2}$ has been widely investigated by the theoretical calculations at the absence of SOC, and
semiconductor-metal phase transition has been predicted by using both tensile and compressive strain\cite{q13,q131,q132}.
Here, we investigate the biaxial strain dependence of electronic structures and thermoelectric properties
by using LDA+SOC.  The energy band gap  and   spin-orbit splitting value at VBM  as a function of $a$ by using LDA+SOC  are present in \autoref{t3}, and the energy band structures for considered $a$ are also displayed  in \autoref{t4}. The energy band gap  firstly increases, and then decreases with increasing $a$. The corresponding strain changes from compressive one to tensile one. Both compressive and tensile strain can induce the direct-indirect-gap crossover, due to changing from one point of $\Gamma$-K line  to K point for CBM and from K to  $\Gamma$ for VBM with increasing $a$.
The spin-orbit splitting monotonically increases with the increasing $a$, but  has little change about 0.02 eV with $a$ varying  from 3.03 $\mathrm{{\AA}}$  to 3.23 $\mathrm{{\AA}}$.

\begin{figure}
  \includegraphics[width=7cm]{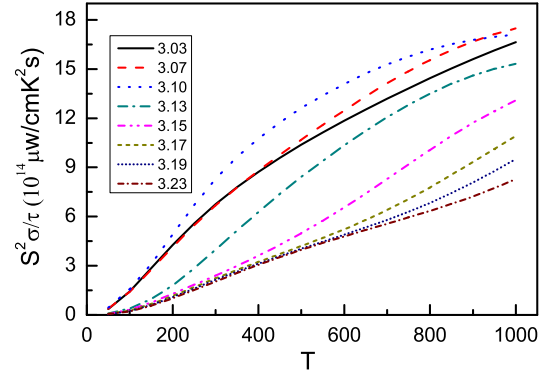}
  \caption{(Color online) Power factor with respect to scattering time $\mathrm{S^2\sigma/\tau}$ as a function of temperature for n-type   with $a$ being from 3.03 $\mathrm{{\AA}}$ to 3.23 $\mathrm{{\AA}}$   calculated by using LDA+SOC  with the doping concentration of $\mathrm{1.23\times10^{13}cm^{-2}}$ (about 0.01 electrons).}\label{t6}
\end{figure}

The strain dependence of S,  $\mathrm{\sigma/\tau}$ and $\mathrm{S^2\sigma/\tau}$   with $a$ changing  from 3.03 $\mathrm{{\AA}}$  to 3.23 $\mathrm{{\AA}}$ calculated by using LDA+SOC at temperature of 300 K are plotted in \autoref{t5}.
The complex  dependence  of strain is observed, due to the sensitive dependence of energy band structures on the applied strain.
By analysing the energy band structure and the corresponding power factor, strain driven  accidental degeneracies can explain strain dependence of power factor.
In n-type doping, the larger S can be attained, leading to larger power factor, when the energy level of  some  conduction band extrema is closer.
For these calculated $a$, the largest S and  $\mathrm{S^2\sigma/\tau}$ can be attained with $a$=3.10  $\mathrm{{\AA}}$ in n-type doping due to the near degeneracy between conduction band extremum along $\Gamma$-K line  and one at  K point.
The same mechanism can be used for p-type, and when
the energy level of some valence peaks is more adjacent, the greater power factor can be gained. For p-type,  S and  $\mathrm{S^2\sigma/\tau}$ reach the peak with $a$=3.17  $\mathrm{{\AA}}$, because the  near degeneracy happen to be induced between $\Gamma$ point and K point near the Fermi level in the valence bands. When  compressive  strain is applied, the direct-indirect gap transition is induced, and the corresponding critical $a$ can produce the largest power factor for n-type in the considered $a$ and doping range.   The  tensile strain can lead to the greatest power factor for p-type at the critical $a$ of direct-indirect gap transition. It is found that $\mathrm{MoS_2}$ has larger power factor in n-type doping than in p-type doping by using compressive strain.
Here, the $\mathrm{S^2\sigma/\tau}$ as a function of temperature  with the doping concentration of  $\mathrm{1.23\times10^{13}cm^{-2}}$ for n-type are plotted in \autoref{t6}.
In the wide temperature range, the power factor with $a$=3.10  $\mathrm{{\AA}}$ is the largest among the considered $a$.

\section{Discussions and Conclusion}
The SOC can  remove the band degeneracy,  which produces remarkable influence on the power factor. The SOC can lead to  detrimental influence on power factor in $\mathrm{Mg_2Sn}$\cite{so1} and half-Heusler $\mathrm{ANiB}$ (A=Ti, Hf, Sc, Y; B=Sn, Sb, Bi)\cite{so2}, especially for p-type doping. For monolayer $\mathrm{MoS_2}$ with optimized lattice constant, SOC not only can reduce power factor in p-type doping, but can enhance one in n-type doping.  When SOC is not included,  the p-type and n-type doping have the near same power factor. However, at the presence of  SOC , the n-type doping shows  more excellent power factor than p-type doping. So, it is very crucial  for power factor calculations  to consider SOC for monolayer $\mathrm{MoS_2}$ in both p-type and n-type doping.

Strain or pressure is a very effective way to realize novel phenomenon by tuning the electronic structures, such as pressure-induced high-Tc superconductivity\cite{ht1,ht2} and strain-induced topological insulator\cite{tps1}. The sensitive strain dependence of electronic structures of  monolayer $\mathrm{MoS_2}$ provides a platform to tune its thermoelectric properties.
The high power factor can be attained by symmetry driven degeneracy, low-dimensional electronic structures and accidental degeneracies\cite{mec}.
Here, the accidental degeneracies  can be induced by both compressive and tensile strain at the critical strain of direct-indirect gap transition, which leads to the larger power factor in certain doping range. Similar pressure induced accidental degeneracies, leading to large power factor, can be found in $\mathrm{Mg_2Sn}$ at the critical pressure of energy band gap\cite{gsd3}.

As is well known, the power factor depends on  the electronic energy structures.  The  electronic structures of  monolayer $\mathrm{MoS_2}$
can not only be tuned  by strain, but by   electric field. In Ref.\cite{q15}, the effect of vertical electric field on the electronic structure of MoS2 bilayer is systematically studied by the first-principle calculations, and the energy  band gap  monotonically decrease with the electric field increasing, leading to the semiconductor-to-metal transition.
Therefore, it is possible to realize improved power factor by applied  electric field.

 In summary,  we investigate strain dependence of  thermoelectric properties  of monolayer $\mathrm{MoS_2}$ by using LDA+SOC, based mainly on
the reliable first-principle calculations.
It is found that including SOC is very important to attain reliable power factor, due to obvious effects of SOC on the energy band structures of monolayer $\mathrm{MoS_2}$.
Calculated results show that strain can realize enhanced power factor at the critical strain of direct-indirect gap transition.
By choosing the  appropriate doping concentration, monolayer $\mathrm{MoS_2}$ under compressive strain in n-type doping  can provide great opportunities for efficient thermoelectricity.

\begin{acknowledgments}
This work is supported by the National Natural Science Foundation of China (Grant No. 11404391). We are grateful to the Advanced Analysis and Computation Center of CUMT for the award of CPU hours to accomplish this work.
\end{acknowledgments}


\begin{references}

\bibitem{s1} Y. Pei, X. Shi, A. LaLonde, H. Wang, L. Chen and G. J. Snyder, Nature \textbf{473}, 66 (2011).

\bibitem{s2} A. D. LaLonde, Y. Pei, H. Wang and G. J. Snyder, Mater. Today \textbf{14}, 526 (2011).
\bibitem{s3} W. S. Liu, Q. Y. Zhang, Y. C. Lan, S. Chen, X. Yan, Q. Zhang, H. Wang,
D. Z. Wang, G. Chen and Z. F. Ren, Adv. Energy Mater. \textbf{1},  577 (2011).

\bibitem{s4}D. K. Ko, Y. J. Kang and C. B. Murray, Nano Lett.,  \textbf{11}, 2841 (2011).



\bibitem{s7}Y. Z. Pei, X. Y. Shi, A. Lalonde  et al, Nature \textbf{473}, 66  (2011).

\bibitem{s8}J. Q. He, J. R. Sootsman, S. N. Girard  et al,  J. Am. Chem. Soc. \textbf{132}, 8669  (2010).


\bibitem{s9}J. F. Li, W. S.  Liu,  L. D. Zhao and M.  Zhou,  NPG Asia Mater.  \textbf{2},
152 (2010).


\bibitem{s10}M. G. Kanatzidis,  Chem. Mater.  \textbf{22}, 648 (2010).


\bibitem{s11}G.  Zhang, B.  Kirk,  L. A.  Jauregui, H.  Yang, X.  Xu,  Y. P. Chen and Y. Wu,  Nano Lett.  \textbf{12}, 56 (2012).



\bibitem{s12}R. Fei,  A. Faghaninia, R.  Soklaski, J. A. Yan,  C.  Lo and L.  Yang,
 Nano Lett.  \textbf{14}, 6393 (2014).


\bibitem{s13}K.  Yang, S.  Cahangirov, A.  Cantarero, A.  Rubio and R. D'Agosta,
 Phys. Rev. B  \textbf{89}, 125403 (2014).






\bibitem{q1}K. F. Mak, C. Lee,  J.  Hone,  J. Shan, and T. F. Heinz,  Phys. Rev. Lett. \textbf{105}, 136805 (2010).


\bibitem{q2}A.  Splendiani, L. Sun, Y.  Zhang, T. Li, J. Kim,  C. Y. Chim,  G.  Galli and  F.  Wang,  Nano Lett.  \textbf{10}, 1271 (2010).


\bibitem{q3} S. W. Han, H.  Kwon,  et al.  Phys. Rev. B  \textbf{84}, 045409 (2011).


\bibitem{q4}S.  Lebegue and O. Eriksson,   Phys. Rev. B  \textbf{79}, 115409 (2009).


\bibitem{q5}C. G. Lee, H. G. Yan, L. E. Brus, T. F. Heinz, J. Hone and S. Ryu, ACS Nano, \textbf{4}, 2695 (2010).


\bibitem{q6}C.  Ataca, H.  Sahin,  E.  Akturk and S. Ciraci,   J. Phys. Chem. C  \textbf{115}, 3934 (2011).

\bibitem{f1}  X. D. Li, J. T. Mullen, Z. H. Jin, K. M. Borysenko, M. B. Nardelli and K. W. Kim, Phys. Rev. B \textbf{87}, 115418 (2013).



\bibitem{q7}S. Ghatak, A. N.  Pal and A.  Ghosh,  Acs Nano \textbf{5}, 7707 (2011).


\bibitem{q8}D. J. Late, B.  Liu, H. R.  Matte, V. P.  Dravid, C. N. R.  Rao,   Acs Nano \textbf{6}, 5635 (2012).


\bibitem{q9}H.  Qiu et al. Appl. Phys. Lett. \textbf{100}
, 123104 (2012).

\bibitem{q10}B. Radisavljevic, A. Radenovic, J. Brivio,	V. Giacometti	and A. Kis, Nature Nanotechnology \textbf{6}, 147 (2011).


\bibitem{q11}X. Zong  et al.  J. Am. Chem. Soc. \textbf{130}, 7176 (2008).




\bibitem{q12}S.  Bhattacharyya  and A. K. Singh, Phys. Rev. B \textbf{86}, 075454 (2012).

\bibitem{q13}E. Scalise, M. Houssa, G. Pourtois, V. Afanas'ev  and A. Stesmans,   Nano Res. \textbf{5}, 43 (2012).
\bibitem{q131}H. Peelaers and C. G. Van de Walle,   Phys. Rev. B \textbf{86}, 241401(R) (2012).


\bibitem{q132}W. S. Yun, S. W. Han, S.  C. Hong, I. G. Kim  and J. D. Lee,  Phys. Rev. B \textbf{85}, 033305 (2012).


\bibitem{q14}A.  Ramasubramaniam, D.  Naveh  and E. Towe,    Phys. Rev. B \textbf{84}, 205325 (2011).


\bibitem{q15}Q.  Liu, L. Li, Y. Li, Z. Gao, Z.  Chen  and  J. Lu,  J. Phys. Chem. C \textbf{116}, 21556 (2012).

\bibitem{t1}H. H Guo, T. Yang,  P.  Tao,  Y.  Wang  and Z. D.  Zhang, J. Appl. Phys. \textbf{113}, 013709 (2013).



\bibitem{t2}S. Bhattacharyya,T. Pandey and A. K. Singh,  Nanotechnology \textbf{25}, 465701 (2014).




\bibitem{t3}K. X. Chen, X. M. Wang, D. C.  Mo and S. S. Lyu,  J. Phys. Chem. C 2015, \textbf{119}, 26706 (2015).

\bibitem{t4} M.  Tahir and U. Schwingenschl\"{o}gl, New Journal of Physics \textbf{16},  115003 (2014).


\bibitem{t5}A. Arab  and  Q.  Li, Sci. Rep. \textbf{5}, 13706 (2015).



\bibitem{so1}K. Kutorasinski, B. Wiendlocha, J. Tobola and S. Kaprzyk,  Phys. Rev. B \textbf{89}, 115205 (2014).

\bibitem{so2}S. D. Guo, J. Alloy. Compd. \textbf{663}, 128 (2016).

\bibitem{gsd3}S. D. Guo, arXiv:1601.02079.

\bibitem{th2}H. Y. Lv, W. J. Lu, D. F. Shao and Y. P. Sun, Phys. Rev. B \textbf{90}, 085433 (2014).

\bibitem{th3}S. V. Ovsyannikov and V. V. Shchennikov, Appl. Phys. Lett. \textbf{90}, 122103 (2007).

\bibitem{th4} S. V. Ovsyannikov, V. V. Shchennikov, G. V. Vorontsov, A. Y. Manakov, A. Y. Likhacheva and V. A. Kulbachinskii, J. Appl. Phys. \textbf{104}, 053713 (2008).
\bibitem{1}P. Hohenberg and W. Kohn, Phys. Rev. \textbf{136},
B864 (1964); W. Kohn and L. J. Sham, Phys. Rev. \textbf{140},
A1133 (1965).

\bibitem{2}P. Blaha, K. Schwarz, G. K. H. Madsen, D. Kvasnicka
 and J. Luitz, WIEN2k, an Augmented Plane Wave
+ Local Orbitals Program for Calculating Crystal Properties
(Karlheinz Schwarz Technische Universit\"at Wien, Austria) 2001,
ISBN 3-9501031-1-2


\bibitem{10}A. H. MacDonald, W. E. Pickett and D. D. Koelling, J. Phys. C \textbf{13}, 2675 (1980).

\bibitem{11}D. J. Singh and L. Nordstrom, Plane Waves, Pseudopotentials and the LAPW
Method, 2nd Edition (Springer, New York, 2006).

\bibitem{12}J. Kunes, P. Novak, R. Schmid, P. Blaha and
K. Schwarz, Phys. Rev. B \textbf{64}, 153102 (2001).

\bibitem{so}D. D. Koelling, B. N. Harmon, J. Phys. C Solid State Phys.  \textbf{10}, 3107 (1977).



\bibitem{b}G. K. H. Madsen and D. J. Singh, Comput. Phys. Commun. \textbf{175}, 67
(2006).

\bibitem{b1}B. L. Huang and M. Kaviany, Phys. Rev. B \textbf{77}, 125209 (2008).

\bibitem{b2}L. Q. Xu, Y. P. Zheng and J. C. Zheng, Phys. Rev. B \textbf{82}, 195102 (2010).

\bibitem{b3}J. J. Pulikkotil, D. J. Singh, S. Auluck, M. Saravanan, D. K. Misra, A. Dhar and R. C. Budhani,
Phys. Rev. B \textbf{86}, 155204 (2012).



\bibitem{ht1}D. F. Duan, Y. X.  Liu, F. B. Tian, D. Li, X. L. Huang, Z. L. Zhao, H. Y.  Yu, B. B. Liu, W. J. Tian  and  T. Cui,
Sci. Rep.  \textbf{4}, 6968  (2014).

\bibitem{ht2}A. P. Drozdov,	 M. I. Eremets,	 I. A. Troyan,	 V. Ksenofontov	 and  S. I. Shylin, Nature \textbf{525}, 73 (2015).

\bibitem{tps1}W. L. Liu, X. Y. Peng, C. Tang, L. Z. Sun, K. W. Zhang, and J. X. Zhong, Phys. Rev. B \textbf{84}, 245105 (2011).

\bibitem{mec}K. F. Garrity, arXiv:1601.01622.



\end{references}
\end{document}